# Interfacially driven transport in narrow channels

Patrice BACCHIN[a,*]

[a]Laboratoire de Génie Chimique, Université de Toulouse, CNRS, INPT, UPS, Toulouse, France
*corresponding author: patrice.bacchin@univ-tlse3.fr

Abstract

When colloids flow in a narrow channel, the transport efficiency is controlled by the non-equilibrium interplay between colloid-wall interactions and hydrodynamics. In this paper, a general, unifying description of colloidal dispersion flow in a confined system is proposed. A momentum and mass balance founded framework implementing the colloid-interface interactions is introduced. The framework allows us to depict how interfacial forces drive the particles and the liquid flows. The interfacially driven flow (osmotic or Marangoni flows for repulsive or attractive colloid-wall interactions respectively) can be directly simulated in two-dimensional domains. The ability of the model to describe the physics of transport in a narrow channel is discussed in detail. The hydrodynamic nature of osmosis and the associated counter-pressure are mechanically related to the colloid-interface interactions. The simulation shows an unexpected transition from axial plug to pillar accumulation for colloidal accumulation at a channel bottleneck. This transition has important consequences in transport efficiencies. Existing limiting cases, such as diffusio-osmosis, are recovered from the simulations, showing that the framework is physically well-founded. The model generalizes the existing approaches and proves the hydrodynamic character of osmosis, which cannot be fully described by purely thermodynamic considerations.

## 1 Introduction

The transport of colloids inside narrow channels is not only an interesting scientific question but also a challenge in many processes in various fields of application. Flow through pores is a common process in living bodies (kidneys, membrane cells, etc.), in natural systems (aquifers) and in industrial applications (filtration, desalting, etc.) [1]. Beyond these applications, the recent development of microfluidic experiments and the nano-scale engineering of interfaces have revived the question of the effect of colloid-interface interactions on transport in confined channels and through small orifices [2,3].

The impact of physicochemical parameters on transport properties needs to be better understood if industrial or natural applications are to be improved but the problem of a ternary system (colloids, liquid and interface) with a strong exchange of forces by interactions and friction is rather complex. When colloids flow in a narrow channel, non-equilibrium interplay between colloid-interface interactions and hydrodynamics controls the transport efficiencies: i) the mass transport efficiency, which governs the colloid transmission through the channels and ii) the mixture (colloids and liquid) transport efficiency, which determines the energy required for the liquid to flow through the channel.

A way to progress toward a better understanding of this complex interplay is to develop model and simulation tools that unravel the mechanisms taking place in such a process. The transport of colloids in narrow channels involves non equilibrium forces balance and exchanges of forces between the colloids and the solvent molecules. Such entanglement can be described at different levels with different simulation methods (Table 1). At the molecular level, dynamic molecular simulations [4] can describe the effect of interaction forces between molecules on the dynamic transport properties. However, colloid particles are much bigger than solvent molecules and, for this reason, dynamic simulation cannot be applied to a large system involving a strong colloid concentration gradient. The coupling between resolved colloidal particles and fluid flow can be modelled with a mesoscopic approach considering the solvent molecules as a continuum phase: a lattice Boltzmann (LB) method [5,6], dynamical density functional theory (DDFT) [7] or a force coupling method (FCM) [8,9]. In the last decade, these "dynamical forces" methods have allowed fast progress to be made in the understanding of the role played by colloidal interaction on the flow at local scale but they are still limited to several hundred particles. In contrast, the thermodynamical approach (for example, the non-equilibrium thermodynamics approach of Kedem and Katchalsky [10]) allows a large number of particles to be processed statistically; the colloid and the fluid phase are both treated as continuum phases. This type of approach is then classically used to describe osmosis as driven by a gradient of water chemical potential. But osmosis is a far-from-equilibrium phenomenon, so an approach based on local equilibrium thermodynamics is not always valid. Furthermore, such an approach cannot take dissipation via local fluid flow into account. Consequently, a complete theoretical description of flow driven by hydrodynamics and several thousands of multi-body interfacial interactions is still lacking.

From a soft matter point of view, progress has recently been made in modelling the role played by colloid-colloid interactions but a better understanding of interfacial phenomena could be achieved by considering the mechanical role played by colloid-interface interactions. This role was qualitatively pointed out in the earlier works on Brownian diffusion at the interface [11–13]. In the 1970's, several authors [14,15] put forward theoretical models to quantify the role of colloid-interface interactions on the transport. However, these models were not really developed and integrated in simulation code. In the past ten years, several authors have again pointed out the important effects of solute or colloid-wall interactions on transport through a narrow channel [16–19]. Recently, a two-fluid model has been proposed that introduces nanometric scale colloid-interface interactions in momentum and mass balances [20]. The two-fluid model or suspension balance model [21,22] is based on solving the field equations written from the volume averaging of the governing equations (local momentum and mass balances) on the two phases: colloids and solvent molecules (Table 1). A 1D application of the model proves that the colloid-interface interaction drives the osmotic flow [23] and, for limiting cases, underlines the compatibility of this mechanical approach with NET approaches. Such models can help to elucidate the "strange" transport mechanism of fluids at the nano-scale [24] (recently highlighted in microfluidic experiments or with nanotubes, aquaporins, etc.) and therefore to help progress by designing specific nanoscale molecule/pore interactions within artificial nano-pores in order to optimize the transport [25].

The aim of this paper is to use the suspension balance model to analyse, with 2D simulations, the coupling of the colloid-wall interaction and hydrodynamic forces when colloidal dispersions are flowing through a narrow channel constriction. The paper is organized as follows. Section 2 introduces the numerical model. In section 3, a limiting case, for a closed channel, is first investigated to analyse a pure interfacially driven transport controlled by a diffusion-osmosis phenomenon [17]. In section 4, the flow of a colloidal dispersion through a channel is simulated to explore the strong coupling between the flow and the colloid-wall interaction and their consequences on transport efficiencies.

## 2 Theoretical background and model development

The model is based on a two-fluid model. The two-fluid model (or mixture model or suspension balance model) [22,26,27] allows the velocities of the colloid phase, $\boldsymbol{u}_c$, of the fluid phase, $\boldsymbol{u}_f$ and of the mixture phase, $\boldsymbol{u}_m$ (coming from volume averaging, $\phi \boldsymbol{u}_c + (1 - \phi)\boldsymbol{u}_f$ ) to be determined, together with the volume fraction of the colloid phase, $\phi$. Solving it thus relies on the application of the momentum and mass balances written for the colloid phase, for the fluid phase and, by addition, for the mixture phase. In a previous paper, such a two fluid model was adapted to implement the colloid/colloid interactions (via the osmotic pressure) and the colloid/interface interactions (via an energy map) [20]. The following sections establish the model (2.1), present the consequence of the balance between pressure and colloid-interface interactions (2.2) and describe the simulation conditions used in the paper (2.3).

## 2.1 Transport of colloids and fluid near interfaces

When colloids flow close to an interface, multi-body interactions occur between colloidal particles and between the colloids and the interface. Classically, colloid-colloid interactions are accounted for by the osmotic pressure, $\Pi_{cc}$, which is a pressure (an energy per unit volume) that includes the entropic contribution and multibody interactions. The gradient of the osmotic pressure leads to a thermodynamic force. This force is responsible for the Brownian diffusion (for the entropic part of the force) and for an interaction-induced diffusion (for the colloid-colloid contribution), both of which are represented by the generalized Stokes Einstein law. On the other hand, the colloid/interface interactions can be defined as an energy required to access a spatial position close to the interface. These interactions can be represented by an energy landscape, $\Pi_{ic}$, that can be spatially mapped. Physically, this landscape [28] maps the colloid-interface interaction energy per unit volume that expresses the Gibbs free energy changes caused by the interactions (similarly to the $\Delta G$ caused by a reaction), which is also the additional free energy for the introduction of a colloid inside the interfacial layer. This map represents the overall interactions between the colloids and the interface (for example, DLVO and hydration forces [29,30]). This colloid/interface interaction energy per unit volume, $\Pi_{ic}$, is also a pressure and is thus a complement to the osmotic pressure accounting for the colloid/colloid interactions, $\Pi_{cc}$. The gradient of the energy map, $\nabla \Pi_{ic}$, gives the force experienced by the colloid close to the interface. The product of the colloid volume fraction by the force, $\phi \nabla \Pi_{ic}$, represents the density of the interfacial force between the colloids and the interface in newtons per unit volume and can be integrated in mechanical approaches [31,32]. Summarizing, the descriptors of colloidal forces are:

- for colloid/colloid forces, the gradient of osmotic pressure, $\nabla \Pi_{cc}$, where $\Pi_{cc}$ is a colligative property (a function of the volume fraction of the dispersion, $\phi$)
- for colloid/interface forces, the density of colloid-interface forces, $\phi \nabla \Pi_{ic}$, where $\Pi_{ic}$ is a map of energy per unit volume (a function of the spatial positions, x, y and z).

The suspension balance model consists of momentum and mass balances written for the fluid, the colloids and the mixture phases (Eqs. 1-6). The delicate part of the model is the writing of the momentum exchanges between the phases that govern the rheological response of the system and the coupling between colloidal and hydrodynamic forces. A physics-grounded expression thus considers that [20]:

- for colloid/colloid forces, a reversible exchange of momentum, $\nabla \Pi_{cc}$, operates between the fluid (Eq. 1) and the colloid phase (Eq. 2) to be consistent with the fluctuation-dissipation theorem for Brownian objects. The colloid/colloid forces thus do not appear in the mixture balance (Eq. 3).
- for colloid/interface forces, by the action/reaction principle, the momentum due to the force, $\phi \nabla \Pi_{ic}$, in the colloid phase (Eq. 1) is counterbalanced by a force applied to the interface. In this case, Newton's third law is broken for the colloid/fluid mixture. There is thus no counteracting force on the fluid (Eq. 2). For these reasons, the colloid/interface force appears in the mixture balance (Eq. 3).

With these considerations, the set of Eulerian equations is given below for a colloidal dispersion with a volume fraction, $\phi$, that corresponds to a particle number density, $n$.

Momentum balance

On the dispersed phase

$$+ n\boldsymbol{F}_{drag} \quad - \nabla \Pi_{cc} - \phi \nabla \Pi_{ic} = 0 \tag{1}$$

On the fluid

$$- n\boldsymbol{F}_{drag} - \nabla p + \eta_m \nabla^2 \boldsymbol{u}_m + \nabla \Pi_{cc} = 0 \tag{2}$$

On the mixture

$$-\nabla p + \eta_m \nabla^2 \boldsymbol{u}_m - \phi \nabla \Pi_{ic} = 0 \tag{3}$$

Mass balance
On the dispersed phase

$$\frac{\partial \phi}{\partial t} = -\nabla \cdot (\phi \boldsymbol{u}_c) \tag{4}$$

On the fluid

$$\frac{\partial (1-\phi)}{\partial t} = -\nabla \cdot \left((1-\phi)\boldsymbol{u}_f\right) \tag{5}$$

On the mixture

$$0 = \nabla \cdot \boldsymbol{u}_m \tag{6}$$

For the drag force representing the forces due to the friction induced by the relative velocity between the phases and the colloid mobility, $m$:

$$\boldsymbol{F}_{drag} = \frac{u_m - u_c}{m(\phi)} \tag{7}$$

By combining Eqs (1-7), a final set of three equations can be obtained:

$$\nabla \cdot \boldsymbol{u}_m = 0 \tag{8}$$

$$-\eta_m \nabla^2 \boldsymbol{u}_m + \nabla p + \phi \nabla \Pi_{ic} = 0 \tag{9}$$

$$\frac{\partial \phi}{\partial t} = -\nabla \cdot \left(\phi \boldsymbol{u}_m + m(\phi) V_p (-\nabla \Pi_{cc} - \phi \nabla \Pi_{ic})\right) \tag{10}$$

where Eq. 8 expresses the conservation of the volume for an incompressible mixture consisting of fluid and colloid; Eq. 9 describes the viscous dissipation of the flow, $-\eta_m \nabla^2 \boldsymbol{u}_m$, driven by the release of fluid pressure, $\nabla p$, and of colloid-interface interactions, $\phi \nabla \Pi_{ic}$; and Eq. 10 relates the colloid flow driven by the mixture flow, the colloid-colloid interaction and the colloid-interface interactions.

The coupling of the mass flux terms (Eq. 10) helps to describe the main mechanisms that occur during transport of colloids through a narrow channel:

- the coupling of convective, $\phi \boldsymbol{u}_m$, and diffusive fluxes, $-mV_p \nabla \Pi_{cc}$, describes the accumulation of concentration [33,34] that takes place when colloids accumulate at the channel bottleneck
- the coupling of convective flux and migration induced by colloid-interface interactions, $-mV_p \phi \nabla \Pi_{ic}$, describes the heterogeneous critical flux phenomena [35,36]: a critical convective drag force needed to overcome the colloid-interface repulsion and then to lead to a deposit at the interface (heterogeneous liquid-solid transition)
- the coupling of diffusive flux and migration induced by colloid-interface interaction describes the Boltzmann exclusion at equilibrium that is due to the colloid-interface interaction [23]

However, a strong difference with conventional approaches can be seen in the momentum balance (Eq. 9). The viscous dissipation term is expressed here as the combination of the applied static pressure with the density of colloid-interface forces, $\phi \nabla \Pi_{ic}$, as previously discussed by Anderson [31]. This coupling allows consistency to be maintained for the description of the equilibrium: in the absence of drag force (i.e. at equilibrium, $\boldsymbol{u}_m = \boldsymbol{u}_c$ in Eq. 2), Eq. 10 leads to $\phi \nabla \Pi_{ic} = -\nabla \Pi_{cc}$ and the momentum equation (Eq. 9) thus matches the description of the equilibrium between the static pressure and the osmotic pressure, $\nabla p = -\phi \nabla \Pi_{ic} = \nabla \Pi_{cc}$. In non-equilibrium conditions, the density of colloid interface forces plays the role of a forcing term (or a local resistance) on the momentum equations of the fluid flow, similarly to the ones appearing in the Force Coupling Method used to describe multi-phase flows [8,9,37]. It will be shown in the next sections that this coupling allows the interfacially driven transport to be described as diffusio-osmosis and Marangoni flows.

## 2.2 Interfacial pressure and interfacially driven transport

The key point discussed in this paper is the coupling of the near-wall colloid and solvent transport phenomena. The coupling between mass and momentum balances is classically taken into account by the viscosity function that encompasses the fluid/colloid interactions. Here, such coupling also comes from the colloid-interface interactions that appear in the momentum (Eq. 9) and mass balance (Eq.10) with the mathematical contribution, $\phi \nabla \Pi_{ic}$ (the interfacial force density in N/m$^3$):

- In the mass balance, these interactions induce an additional mass transport due to the forces between the colloids and the wall
- In the momentum, the interaction with the interface changes the momentum of the colloids and thus contributes to a change in the mixture momentum

The mixture momentum (Eq. 9) is a balance between a viscous dissipation term and two elastic terms: the pressure gradient and the term coming from the colloid-interface interaction, $\phi \nabla \Pi_{ic}$. For greater clarity, the balance between the elastic contributions, $\nabla p = -\phi \nabla \Pi_{ic}$, can be considered first. A change in colloid-interface interactions can thus be associated with a fluid pressure change. If the colloid-interface interactions are repulsive, ($\Pi_{ic}$ increases when approaching the interface), these interactions lead to a decrease in the fluid pressure close to the interface. Physically, this means that the fluid loses the mechanical reaction force (resulting from the action force of the colloid on the interface), which is dissipated in the interface (in absence of elastocapillary effect) [23]. This decrease in fluid pressure at the interface allows the total pressure applied on the interface to be kept constant: the loss in fluid pressure is equated to the pressure due to the normal force exerted by colloids on the interface. This contribution of the colloid-interface interaction to the fluid pressure can be computed as the integration of the force density from the bulk to a position A in the system, $p_{int} = -\int_\infty^A \phi d\Pi_{ic}$, the so called "interfacial pressure". Repulsion (attraction) with the interface leads to a negative (positive) interfacial pressure that can be associated with a local decrease (increase) in fluid pressure close to the interface.

Local changes in the interfacial pressure can appear because of variation in the colloid volume fraction, $\phi$, or variation in the colloid-interface interaction, $\Pi_{ic}$. If such variations occur along an interface, a pressure gradient develops along the surface which, in turn, induces a flow called interfacially driven transport. Such phenomena are sketched for the cases of attractive and repulsive colloid-interface interactions in Fig. 2(a) and Fig. 2(b), respectively. In the case of attractive interaction (Fig. 2a), the presence of colloids interacting with the wall induces a liquid overpressure. In this case, the increase in colloid concentration induces an increase of fluid overpressure. A flow is then induced from the high pressure (high concentration) zone to the low concentration zone. This flow corresponds to a Marangoni effect (also called capillary driven flow) that drives the flow to low surface tension (where there is a large number of colloids having an affinity with the interface i.e. acting as a surfactant) to high surface tension (where there are few surfactant-colloids). In the case of repulsive interactions, Fig. 2(b), the interfacial interactions induce a decrease in the interfacial pressure in the region with a high concentration of colloids. This pressure gradient drives a flow called diffusion-osmosis flow in this case.

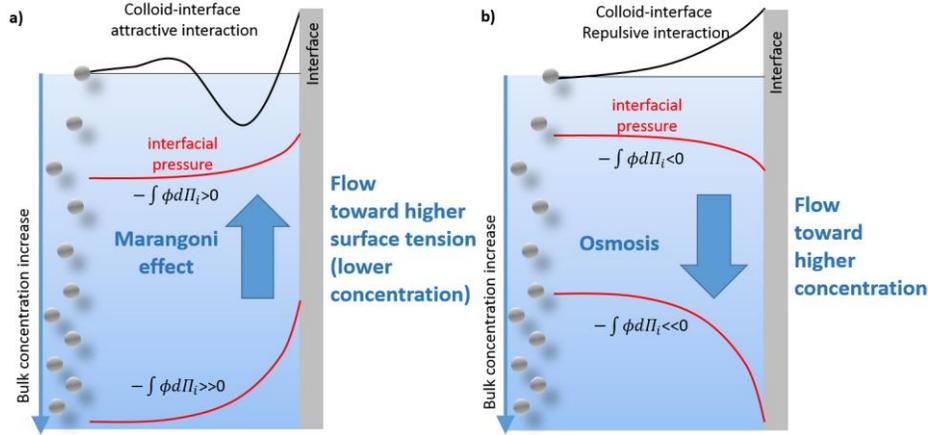

*Fig. 1: Representation of the interfacially driven transport in presence of a gradient of colloid concentration tangentially to an interface. When the colloids are attracted by the interface (a) the colloid-interface interaction contributes to a local increase in the interfacial pressure contribution, $-\int \phi d\Pi_{ic}$. This increase is greater when the colloid concentration is higher (bottom part of the figure). This tangential pressure gradient leads to flow directed toward the lower concentration: Marangoni flow. For colloids experiencing repulsion with the interface b), the interaction leads to a local decrease of pressure close to the interface that is more pronounced for higher colloid concentration. These pressure variations induce a flow toward the higher concentration. This diffusio-osmotic phenomenon causes the osmotic flows.*

The interfacial force density, $\phi \nabla \Pi_{ic}$, drives the flow of both particles (particle exclusion in interfacial layer) and liquid (through interfacially driven flows). These interfacial forces between the colloids and the interface strongly couple the near-wall colloid and solvent dynamics. These transport phenomena thus have a hydrodynamic character that cannot be explained by thermodynamic considerations. The two-fluid model that accounts for colloid-interface interaction in momentum and mass balances depicts the transient initiation of the osmotic and Marangoni flows and thus generalizes and unifies the existing approaches.

This description also has further consequences when the interface is no longer a solid interface but a fluid one. From a thermodynamical point of view, the energy map represents the interfacial Gibbs free energy and, thus, the increase or decrease in pressure is closely related to an increase or a decrease in water activity, $p_{int} = \frac{kT}{V_w}\ln(a_w)$. Considering water activity instead of interfacial pressure, it can be noted in Fig. 1 that the osmotic or the Marangoni flows always develop from high to low water activity zones. Such local variation of water activity at the interface can have many consequences when evaporation kinetics is considered, the evaporation rate being proportional to the difference of water activity between the liquid interface and the ambient air. The local variation of water activity given by the model can thus help to consider the contribution of colloid-colloid and colloid-interface interactions on water evaporation. The interaction of colloids with the interface also leads to an anisotropy in pressure close to the interface. This stress anisotropy can be normal and/or tangential to the interface [38] :

- when there is a local pressure gradient normal to the free fluid surface, it leads to an interfacial (capillary) stress that leads to a surface tension. According to Kirkwood and Buff [39], the surface tension for a flat interface is linked to the integral of the interfacial pressure across the interface, $\gamma = \int_{-\infty}^{\infty} p_{int} dx$.
- when there is a pressure gradient tangential to the surface (because of a variation of colloid volume fraction or a variation in interactions), interfacially driven flows (solvent flows with a component tangential to the surface) are initiated.

## 2.3 Simulation of interfacially driven flow in a narrow channel

In this paper, the set of equations (Eqs. 9-11) is solved in a geometry representing a channel with solid (non-deformable) walls with a non-dimensional form:

$$\nabla . Pe = 0 \qquad (12)$$

$$\nabla . (\hat{\mu}(\phi) \nabla Pe) = \nabla \hat{p} + \phi \nabla \hat{\Pi}_{ic} + (1-\phi) \nabla \hat{\Pi}_{if} \qquad (13)$$

$$\frac{\partial \phi}{\partial \hat{t}} = -\nabla . \left( \phi Pe + K(\phi) \left( -\nabla \hat{\Pi}_{cc} - \phi \nabla \hat{\Pi}_{ic} \right) \right) \qquad (14)$$

In Table 1, the dimensionless variables are defined and their links with the dimensional variables are quantified. The non-dimensional terms are obtained by dividing by the diffusion force so that the advection term becomes a Péclet number in Eqs. 12-14. The corresponding Reynolds number is thus the Péclet number divided by the Schmidt number, Sc. The set of data used in these simulations was defined to characterize a dispersion of 10 nm in diameter in a system with a characteristic size (the pore size) of 1 micrometre. Such a size ratio ensures that the dispersion is treated as a continuous medium and the Eulerian approach is correctly used. Under these size conditions, the non-dimensional viscosity is equal to $5.55 \, 10^{-6}$. The dependence of viscosity on the volume fraction is not taken into account. The osmotic pressure is defined by a van't Hoff law for an ideal dispersion: the colloid-colloid interactions are not taken into account and the diffusion coefficient remains independent of the colloid volume fraction. In these simulations, the mechanisms induced by colloid/colloid interactions and by the coupling between volume fraction and viscosity are switched off. Such simulations thus focus on the effect of colloid/interface interactions on the fluid dynamics. These simulations can represent the flow of a rather dilute colloidal dispersion close to an interacting interface.

Table 1: The dimensionless quantities used to define the dynamic osmotic problem. The correspondence with the dimensional quantities is given for the conditions of $a=10^{-8}$ m, $\delta = 10^{-6}$ m, $\mu = 10^{-3}$ Pa.s, and T=298 K.

| Quantity | Dimensionless form | Correspondence |
|---|---|---|
| **Velocity** | Péclet $Pe = \frac{u_m \delta}{m_0 kT}$<br>Reynolds $Re = Pe/Sc$ | $u \, (m) = 2.18 \, 10^{-6} \, Pe = 0.1 \, Re$ |
| **Viscosity** | $\hat{\mu} = \frac{2a^2}{9\delta^2} \frac{\mu(\phi)}{\mu_0}$ | $\mu(Pa.s) = \mu_0 4.5 \, 10^{+6} \hat{\mu}$ |

| Pressure | $\widehat{p} = \dfrac{V_p p}{kT}$ | $p(Pa) = 982\, \widehat{p}$ |
|---|---|---|
| Time | $\widehat{t} = \dfrac{m_0 kT}{\delta^2} t$ | $t(s) = 4.58\, \widehat{t}$ |
| Osmotic pressure or interfacial pressure | $\widehat{\Pi} = \dfrac{V_p \Pi}{kT}$ | $\Pi(Pa) = 982\, \widehat{\Pi}$ |
| Mobility | Settling hindrance coefficient $K(\phi) = \dfrac{m}{m_0}$ | $m\,(kg^{-1}.s) = 5.31\,10^{+9}\, K(\phi)$ |

Eqs. 12-14 could be solved in specific geometries by introducing solid walls as boundary conditions with no-slipping conditions. In this paper, another approach is chosen: the equations are solved for the whole domain but with a local penalization method [40] in the Stokes equation to account for the presence of solid walls. In Eq. 13, a term, $\widehat{\Pi}_{if}$, is thus added to penalize the flow in the solid domain. This term physically express the fluid-wall interaction that forces the flow away from the interface. This way of writing the equation has the advantage of treating the wall interactions similarly: the presence of the narrow channel in the flow is represented through the interactions that the wall interface exerts on the fluid, $\widehat{\Pi}_{if}$, together with the interaction it exerts on the colloids, $\widehat{\Pi}_{ic}$. These interactions are a function of the distance to the wall, which is determined through a level set method. The penalization for the fluid is a very stiff exponential function that applies in a very thin interfacial layer close to the interface. To be negligible, the interfacial layer for fluid-interface interactions is less than one tenth of the interfacial layer for colloid-interface interactions. The interaction between the colloids and the wall are also represented by an exponentially decreasing function similar to the one that could be obtained by the DLVO theory.

In this paper, the decay length is taken to be 0.1 (one tenth of the pore diameter) and the maximum energy at the wall is fixed at 100. These values were chosen to be close to those calculated for 10 nm spheres dispersed in $10^{-5}$ M solution with zeta potential of 80 mV for both particles and walls. The resulting colloid-interface interaction map is plotted in Fig. 2.

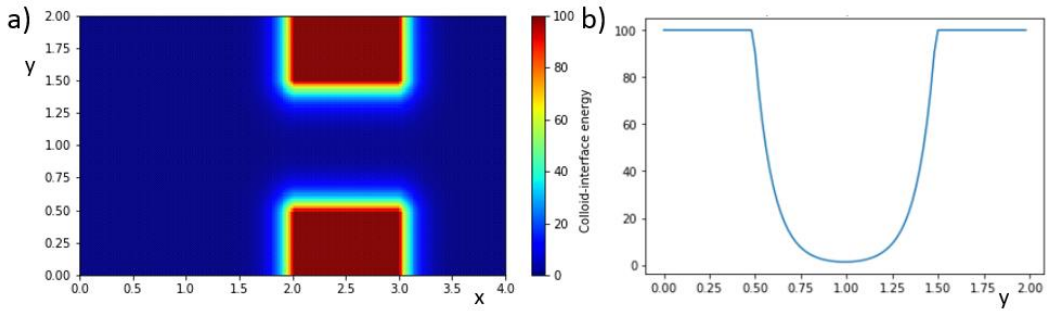

*Fig. 2. a) 2D representation of the narrow channel geometry (the pillars are represented in brown) and of the colloid-wall interaction magnitude (with the colour map) b) Colloid-wall interaction energy across the narrow channel opening for x=2.5.*

The set of equations was solved with the partial differential equation solver Fipy [41] (finite element volume) implemented on the Python platform Canopy (Enthought, Austin). Simulations were performed with periodic conditions on the top and bottom boundaries. The full code used for the solving is given in SI 6. Simulations are presented in the next sections i) for no net flow conditions through the channel to illustrate the diffusion-osmosis phenomena (section 3) and ii) for filtration conditions through the channels with a counter-osmotic pressure (section 4).

# 3   Pure diffusio-osmosis in closed channel

In this section, the model is used to describe a diffusio-osmotic flow by simulating the flow induced by a gradient of concentration. The simulations are performed in a 2D geometry representing a pore closed at one end (right boundary in Fig. 3). The global x velocity is thus zero. At the open side of the channel (left boundary), colloids arrive by diffusion. This diffusion induces a gradient of concentration along the pore axis that, in turn, should lead to an osmotic flow: the so called diffusio-osmosis phenomenon.

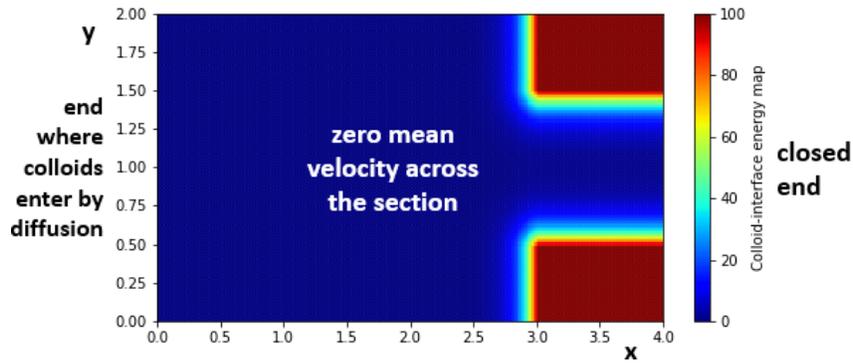

Fig. 3: Representation of the colloid-interface energy map in the semi-closed channel (axis of the channel in the x direction). The boundary conditions are periodic at the top (y=2) and bottom (y=0), wall conditions on the right (zero velocities and zero mass flux at x=4) and a mass inlet with no flow on the left (constant concentration and zero velocities at x=0).

Simulations allow the momentum and the mass balance (Eqs. 13 and 14 respectively) to be solved and the continuity equation (Eq. 12) to be satisfied. The mixture velocities along x and y, the pressure field and the volume fraction field are determined. The transient development of the diffusion-osmosis flow are illustrated with animations showing the spatiotemporal evolution of these fields (SI 1). The simulations describe:

- the generation of an osmotic flow (with negative x velocities) when the colloid concentration reaches the pore wall (diffusion-osmosis)
- the coupling of the osmotic flow with forced convection (with positive x velocities) to keep a zero net flow across the pore section (diffusion-osmotic / forced convection secondary flow)
- the return to equilibrium (zero fluid velocities and zero mass flux) when the diffusion homogenizes the concentration along the pore axis.

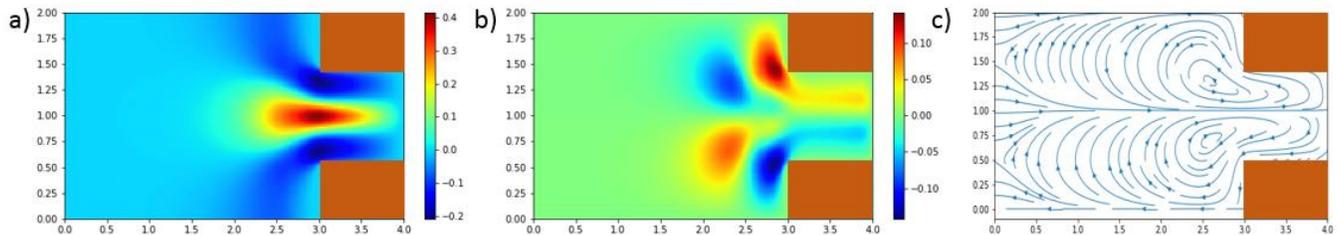

*Fig. 4: Intensities of x velocities and y velocities (represented by a dimensionless Péclet number) in a) and b) respectively, for a time t=2. These variations correspond to two secondary flows with negative x velocities close to the pore walls and a return flow (positive x velocites) in the pore centre. The representation of stream lines in c) highlights the two recirculation regions at the channel inlet.*

Fig. 4 represents the velocities (presented in a non-dimensional form as a Péclet number here) in the x direction (Fig. 4a) and in the y direction (Fig. 4b) at time t=2 (i.e. when the diffusion-osmosis phenomenon is at its maximum). The stream line representation (Fig. 4c) illustrates the presence of the secondary flow with negative x velocities close to the wall (the osmotic flow) and a positive x velocity in the pore centre (the forced convection developing to ensure a zero net flow through the pore). These secondary flow cells have positive and negative y velocities at their ends (Fig. 4b). The flows are characterized by stationary planes where the x velocities are zero. These stationary planes can be seen in Fig. 5, which presents the x velocity profile along y at the pore entrance (x=3). The stationary planes are located at 30 % and 70 % of the pore diameter. These locations correspond to $2^{-0.5}$ and $1 - 2^{-0.5}$ that can be determined from the analytical solution for an interfacial flow combined with forced convection in Cartesian coordinates (as with electro-osmosis flow).

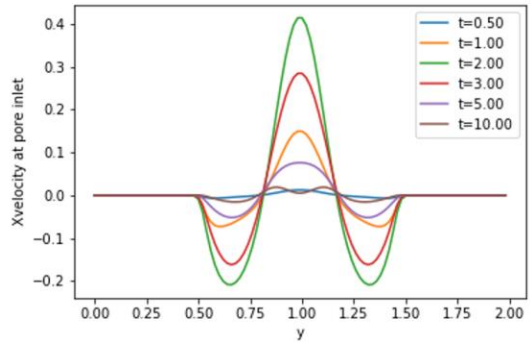

*Fig. 5: x velocity profile along y at the pore entrance (x=3) for different times. The stationary planes are located at 30 % and 70 % of the pore diameter.*

The secondary flows become established progressively with time. Fig. 6 presents the variation with time of the osmotic flow rate defined as the integration along y of the negative x velocity presented in Fig. 6. There is no osmotic fluid flow at the initial time when the colloids are not interacting with the pore wall. As soon as the colloids interact with the walls, the contribution of $\phi \nabla \widehat{\Pi}_{ic}$ in the momentum equation leads to the osmotic flow in the direction of higher $\phi$ value. The maximum flow rate occurs for t=2 and therefore the flow starts to decrease mainly because the concentration gradient decreases along the pore axis. The volume fractions of colloids are given in SI 2. For longer times (i.e. when diffusion homogenizes the concentration) the osmotic flow vanishes.

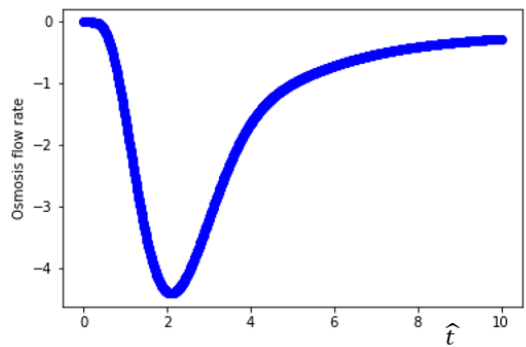

*Fig. 6: Evolution of the half osmotic flow rate (resulting from the integration along y of the negative x velocities at the pore entrance Fig. 5) with time. At the beginning, when colloids are not interacting with the wall, the osmotic flow rate is zero. The osmotic flow is maximum when the colloid concentration gradient reaches the pore wall and vanishes again when the diffusion homogenizes the concentration in the pore.*

To conclude, the diffusio-osmosis test case of a pore with one end closed allows us to check the ability of the model to describe i) how osmotic solvent flow transiently initiates from colloid/interface interactions, ii) how interfacially driven transport and forced convection act together to generate secondary flows, and iii) how the system returns to equilibrium. The diffusion-osmosis mechanism can be described according to the following steps. The diffusion of colloids toward the interface generates a concentration gradient parallel to the pore wall. These concentration gradients lead to a local force density, $\phi \nabla \widehat{\Pi}_{ic}$ (plotted in Supplementary Information 3) that initiates the diffusion-osmotic flows. In the case of a closed system, forced convection becomes established in the opposite direction to satisfy the solvent continuity and, combined with the diffusion-osmotic flow, leads to secondary flows close to the interface.

Similar simulations can be run for colloid-interface attraction instead of repulsion. In this case, the secondary flow appears but in the opposite direction. The liquid flow is then directed toward the zone of low colloid concentration. This flow is a solute-capillary Marangoni flow with a flow toward the zone where colloids having an affinity with the interface are less concentrated (or zone of higher surface tension).

# 4 Osmotic flows in channel during filtration

In this section, simulations are performed with a geometry similar to that used in the previous section but with a flow rate through the channels (the narrow channel geometry is presented in Fig. 1). In filtration conditions, the mass boundary conditions are a constant concentration on the left side (inlet of a flow with a given concentration) and a concentration gradient at zero on the right (outlet flow). The flow boundary conditions in the right and the left domain boundaries are a constant Péclet number along x (fixing the net flux through the pore) and a Péclet number along y at zero. The overall Péclet number along x (also corresponding to a fluid velocity or a Reynolds number as presented in Table 1) is then fixed by the boundary conditions. The boundary conditions are periodic on the top and bottom meshes. This corresponds to a filtration case with a constant flow rate through the narrow channel: the pressure drop thus increases if colloids accumulate at the pore bottleneck and offer resistance to the flow. These simulations then depict the reverse osmosis situation where a counter osmotic flow acts against a forced convection. The previous case, described in section 3, can be seen as the asymptotic case for a Péclet number of zero, where diffusion-osmosis occurs in the absence of a net flow through the channel.

Simulations have been performed for different Péclet numbers between 0.1 and 10. The full data of the transient simulations are given in an animated panel presenting the evolution of the main parameters with time in supplementary materials (SI 4). To analyse the simulation results, the simulation with a Péclet number equal to 3 will be considered first. Fig. 7a maps the volume fraction accumulated at the pore entrance for the final time t=10 (a quasi-steady state is reached at this time) and Fig. 7b presents the variation of the volume fraction along the channel axis with time.

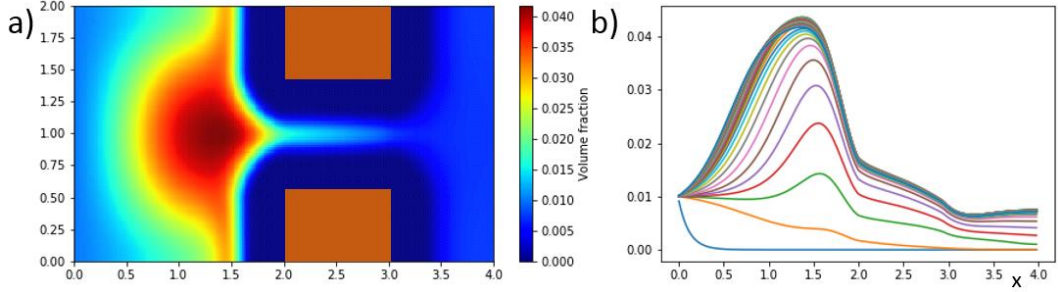

*Fig. 7: Spatiotemporal variation of colloid volume fraction at Pe=3. Fig. 7a maps the volume fraction on the domain at t=10. Fig. 7b presents the volume fraction profile along the channel axis, x (for y=1) for different simulation times with a time step of 0.25 from t=0 to t=10.*

Colloids accumulate because of partial retention due to the repulsive barrier along the wall of the pore channel (Fig. 2a): the water can flow in these regions whereas the colloids are expelled. The accumulation takes the form of a hemispherical plug at the pore bottleneck (Fig. 7a) where the volume fraction of colloids can be five times higher than that coming from the bulk on the left. The plug builds up progressively with time and reaches a steady state shape when the retro-diffusion in the bulk balances the convective mass flux on the left boundary. The thickness of the upstream area thus acts as a mass boundary layer that is progressively filled by the accumulation.

The flow is modified by the accumulation, which leads to a concentration gradient along the channel and then to the diffusio-osmosis phenomenon. This osmotic flow is opposite to the forced convection and appears as negative x velocities in Fig. 8a. These osmotic flows are the direct consequence of the contribution of the colloid-wall interaction (term $\phi \nabla \Pi_i$ in the momentum balance). This interfacial contribution acts as if there was a decrease in pressure close to the wall (where $\nabla \Pi_i$ is large) or in the concentrated zone in the upstream channel zone (where $\phi$ is large)[23]. The map of the contribution of $\phi \nabla \Pi_i$ is given in supplementary materials (SI 5). It should be noted that negative x velocities also appear where the colloid accumulates (around the position (1,1) in Figs 7a and 8a). Such negative flows are due to the strong gradients of colloid concentration in the zone where colloids interact with the interface (Fig. 8a), which induces osmosis.

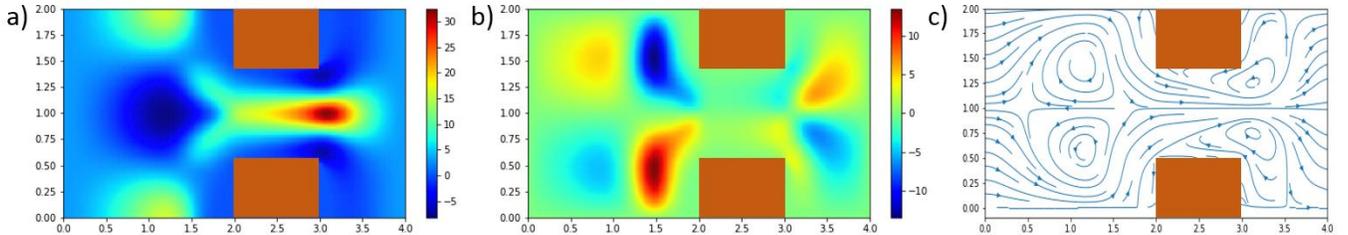

*Fig. 8: Intensities of x velocities and y velocities (represented by dimensionless local Péclet number) in a) and b) respectively for a time t=10 and for a net overall Péclet number along x of 3. These variations correspond to two secondary flows with negative x velocities close to the pore walls and a return flow (positive x velocites) in the pore centre. The stream lines are represented in c).*

The negative osmotic flows result in an additional pressure drop that keeps the permeate flow rate constant across the channel. In Fig. 9, the Péclet number (fixed as a boundary condition) is plotted versus the pressure drop at quasi-steady state (determined by simulations). This plot is a non-dimensional form of the classic plot of the permeate flux versus the transmembrane pressure used to quantify the filtration results. In this kind of plot, the dotted line represents the pressure drop simulated for the solvent water and the horizontal gap between the dotted line and the one obtained when filtering the colloid represents the additional counter osmotic pressure.

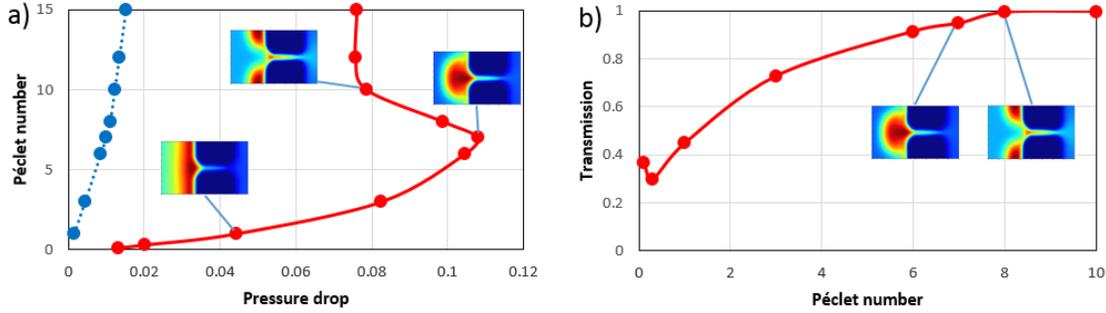

*Fig. 9: a) Péclet number as a function of the pressure drop across the channel (full red line). The dotted line represents the same data when only the solvent is filtered. b) Colloid transmission as a function of the Pe number. The inserts represent the volume fraction map (details given in Fig. 10).*

It can be seen that the counter osmotic pressure does not vary monotonously when the flow rate increases. For a Péclet number around 7, a maximum in the counter osmotic pressure is observed, together with a marked increase in transmission. This change is also associated with a very different pattern for the colloid accumulation as sketched in Fig. 10. The pattern of the accumulation takes different forms:

- a homogeneous accumulated layer when operating at low Péclet number. The diffusion of colloids is high enough to homogenize the concentration along the y direction.
- an axial plug with diffuse accumulated colloids at the channel bottleneck for intermediate Péclet number. This pattern is responsible for the highest accumulation and the highest counter osmotic pressure at a Péclet number of 7.
- an accumulation on the channel pillars with only a partial axial accumulation for larger Péclet numbers. These Péclet numbers also relate to greater transmission through the channel: the flow is able to overcome the colloid-wall interactions.

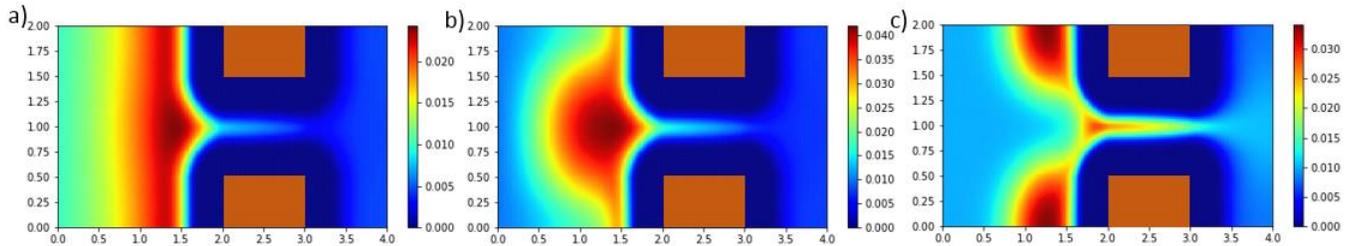

*Fig.10: Colloid accumulation for a) Pe=1, b) Pe=3 and c) Pe=10 for a time t=10 corresponding to a quasi-steady state for all cases. The accumulation patterns are very different with a) a homogeneous layer for low Péclet number (high diffusion that homogenizes the concentration along y) b) an axial plug at the channel bottleneck for intermediate Péclet number and c) an accumulation on the channel pillars with only a partial axial accumulation for high Péclet number.*

The transition between an axial plug (Fig. 10b) and pillar accumulation (Fig. 10c) occurs sharply above the threshold Péclet number of 7. The breaking of the accumulated layer is also observed for the same Péclet number but over time. Fig. 11 presents the way the pillar accumulation pattern is initiated over time for a simulation performed with a Péclet number of 10. Initially, the accumulation grows in the form of a film of homogeneous thickness. With time, a thinning of the accumulation is observed at the edge of the conical bottleneck (Fig. 11b) before a breakage of the accumulation between an axial plug and the pillar accumulation (Fig. 11c). For further time, the axial plug disappears, whereas the pillar accumulation becomes thicker and leads to the steady state accumulation presented in Fig. 10c.

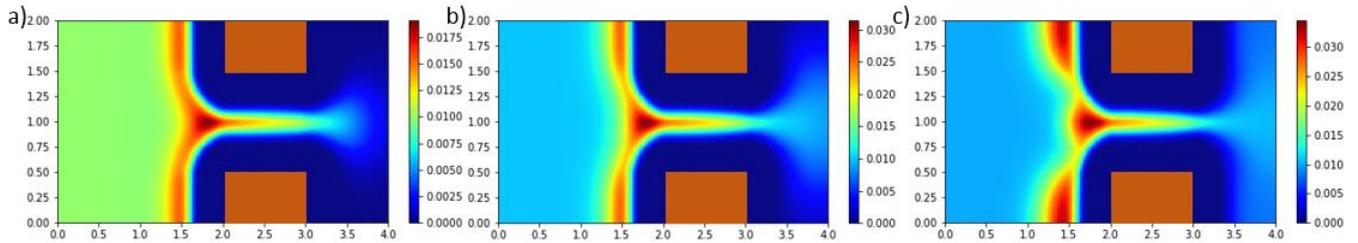

*Fig.11 Dynamic evolution of the accumulation (t=0.2, 0.3, 0.5) for Pe=10. Breakage of the accumulated film to give axial plug and accumulation on channel pillars. The steady state (t=10) accumulation on channel pillars is presented in Fig. 10c.*

These results illustrate how the coupling between diffusion-osmosis and hydrodynamics controls the flow in a channel. A complex force chain operates between the components of the ternary system: interface/colloid/solvent molecules [23]. This force chain can be summarized as follows: i) colloids interact with the interface ii) the colloid/solvent mixture loses momentum due to the reaction force acting on the interface (Newton's third law applies for the ternary interface/colloid/solvent system but is violated for the colloid/solvent pair) iii) for colloidal particles, according to the dissipation fluctuation theorem, this force anisotropy normal to the interface (non-zero transverse stress) leads to a net force density on the fluid near the interface. The three components of the ternary system are then intimately connected by the force chain: at the end, because of colloid/interface interactions, a net force acts on the solvent molecules.

From mechanical balances, the simulations show how this net force modifies the Stokes flow and thus controls the near-wall non-equilibrium dynamics of colloids and solvent molecules. Ultimately, describing the interfacially driven flows (diffusion-osmosis and Marangoni effect) appearing inside the liquid phase is a key point but it is also important, when considering boundary conditions, to depict the surface tension and evaporation kinetics. This local description of the flow dynamics could also help to understand how the pore architecture, e.g. its hourglass shape, changes the transport efficiencies. This kind of model should be useful to depict other conditions where diffusion-osmosis takes place, such as diffusio-phoresis [42] and transport of self-propelling active colloids. Such non-equilibrium conditions could also lead to Rayleigh-Taylor instabilities that could be a way to decrease narrow channel fouling [43].

## 5 Conclusions

In summary, a two phase flow model and 2D simulations allow the dynamics of flowing colloids to be explored when they interact with an interface. From mechanical balances, the colloid/interface interactions lead to a net force acting on the solvent molecules near the interface, which can be ascribed to an interfacial pressure gradient. Interfacially driven transport, like diffusion-osmosis or the Marangoni effect, originates from gradients in interfacial pressure along the interface, for example, due to a gradient in colloid volume fraction. The model explicitly shows how colloid/interface interactions initiate interfacially driven transport. The model thus gives a clear local scale interpretation of the osmosis phenomenon within a mechanical approach. The comprehensive approach generalizes and unifies the description of colloidal dispersion flow in a confined system.

Interesting consequences emerge when the transport of colloids through a narrow channel is considered. Secondary flows develop in the narrow channel bottleneck because of the coupling between diffusion-osmosis and forced flow. The dynamics of these osmosis secondary flows has been analysed here and their consequences on the counter pressure and on the colloid transmission through the channel discussed. These local flows also control the way the colloids accumulate at the channel bottleneck. The simulations shed light on the existence of a critical Péclet number that leads to a transition between axial plug formation and accumulation on channel pillars.

# 6 Acknowledgements

The author thanks Yannick Hallez, Fabien Chauvet, Léo Garcia and Micheline Abbas for fruitful discussions.